\documentclass[pra,twocolumn,showpacs,preprintnumbers,amsmath,amssymb]{revtex4}

\usepackage{graphicx}
\usepackage{epstopdf}
\usepackage{amscd}
\usepackage{dcolumn}
\usepackage{bm}

\newcommand{\metricd}{\operatorname{d}} 

\newcommand{\mment}[1] {\ensuremath{\mathbf{#1}}} 
\newcommand{\map}[1]{\ensuremath{\mathcal{#1}}}

\newcommand{\numberfield}[1]{\ensuremath{\mathbb{#1}}} 

\newcommand{\cvect}[1] {\textsf{#1}} 

\newcommand{\lvect}[1] {\ensuremath{\mathbf{#1}}} 

\newcommand{\cmatrix}[1]{\textsf{#1}} 

\newcommand{\rmatrix}[1]{\ensuremath{#1}} 

\newcommand{\hypersphere}{\ensuremath{\mathit{S}^{2N-1}}} 

\newcommand{\transpose}{\ensuremath{^\text{\textsf{T}}}} 

\addtocounter{secnumdepth}{-1} 

\newcommand{\var}{\ensuremath{\chi}} 

\begin{document}



\title{From Information Geometry to Quantum Theory}
\author{Philip Goyal}
    \email{pgoyal@perimeterinstitute.ca}
    \affiliation{Perimeter Institute, \\ Waterloo, Canada}


\begin{abstract}
In this paper, we show how information geometry, the natural geometry of discrete probability distributions, can be used to derive the quantum formalism.  The derivation rests upon three elementary features of quantum phenomena, namely complementarity, measurement simulability, and global gauge invariance.  When these features are appropriately formalized within an information geometric framework, and combined with a novel information-theoretic principle, the central features of the finite-dimensional quantum formalism can be reconstructed.

\end{abstract}


\maketitle

The unparalleled empirical success of quantum theory strongly suggests that it accurately captures fundamental aspects of the workings of the physical world. The clear articulation of these aspects is of inestimable value not only for the deeper understanding of quantum theory in itself~\cite{Fuchs02}, but for its modification~(for example, to allow non-unitary continuous transformations~\cite{Weinberg89a, Weinberg89b, Herbert82}) and its further development, particularly for the development of a theory of quantum gravity~(see~\cite{Isham-quantum-and-reals}, for example).  However, such articulation has traditionally been hampered by the fact that the quantum formalism, in which these aspects are presumably encoded, consists of postulates expressed in an abstract mathematical language to which our physical intuition cannot directly relate.  Over the last two decades, there has been growing interest in elucidating these aspects by expressing, in a less abstract mathematical language, what quantum theory \emph{might} be telling us about how nature works, and trying to derive, or \emph{reconstruct}, quantum theory on this basis~\cite{Wheeler89, Rovelli96, Popescu-Rohrlich97, Zeilinger99, Fuchs02, Hardy01a}.   

Much of the recent effort in reconstructing the quantum formalism is motivated by the hypothesis that the concept of information might be the key, hitherto missing, ingredient, that may enable a reconstruction, and several attempts have been made to systematically explore the reconstruction of the quantum formalism from an informational starting point~(for example~\cite{Wootters80, Summhammer94, Brukner99, Brukner02b, Rovelli96, Grinbaum03, Grinbaum04, Caticha99b, Clifton-Bub-Halvorson03}).   Although these approaches have yielded significant insights, they are either incomplete~(for example,~\cite{Wootters80, Summhammer94, Brukner02b}) or employ abstract assumptions that involve the assumption of the complex number field~(for example,~\cite{Grinbaum04, Caticha99b,Clifton-Bub-Halvorson03}).  Such assumptions significantly limit the degree to which the physical content of the quantum formalism can be elucidated since one of the most mysterious mathematical features of the quantum formalism is being assumed at the outset.   In this paper, we show that the principal mathematical features of quantum theory can be reconstructed using the concept of information without employing such assumptions.

Our approach develops intimate connections, known to exist for some time, between structures that arise naturally in classical probability theory on the one hand, and the quantum formalism for pure states on the other~\cite{Wootters-statistical-distance, Brody-Hughston96,Brody-Hughston98,Mehrafarin05}. For example, Wootters~\cite{Wootters-statistical-distance} has shown in the framework of classical probability theory that one can quantify the degree to which two discrete probability distributions,~$\lvect{p}=(p_1, \dots, p_N)$ and~$\lvect{p}'=(p'_1, \dots, p'_N)$, can be distinguished given the same number of samples from each by means of the \emph{statistical distance},~$\metricd_S(\lvect{p}, \lvect{p}') = \cos^{-1}\left(\sum_i \sqrt{p_i p'_i}\right)$, between them. If one considers the  statistical distance,~$\metricd_{S}(\lvect{p}, \lvect{p}')$, between the probability distributions~$\lvect{p}$ and~$\lvect{p}'$ which characterize the results of projective measurement~$\mment{A}$ when performed upon two $N$-dimensional pure states~$\cvect{u}$ and~$\cvect{v}$, respectively, and if one chooses~$\mment{A}$ such that~$\metricd_S$ is maximized, Wootters shows that~$\metricd_S$ is equal to the Hilbert space distance,~$\metricd_H(\cvect{u}, \cvect{v}) = \cos^{-1}|\cvect{u}^\dagger\cvect{v}|$, between~$\cvect{u}$ and~$\cvect{v}$~\cite{Wootters-statistical-distance}. The existence of such a connection  is remarkable, and suggest that the usual formalism of quantum theory might owe at least some of its structure to the notion of distinguishability that arises naturally in a purely classical probabilistic setting.

Following Wootters, we adopt an operational approach, and so take the probabilistic nature of measurements as a given.  Accordingly the framework of classical probability theory is taken as a starting point.  We equip this framework with a metric,~$ds^2 = \frac{1}{4} \sum_i dp_i^2/p_i$, the \emph{information metric}~(or Fisher-Rao metric), the infinitesimal form of the statistical distance, rather than the statistical distance itself, as this suffices for the purposes of the reconstruction.  This metric determines the distance between infinitesimally close probability distributions~$\lvect{p} = (p_1, \dots, p_N)$ and~$\lvect{p}' = (p'_1, \dots, p'_N)$.  As we shall describe below, the information metric can be understood as a natural consequence of the introduction of the concept of information into the probabilistic framework.     Accordingly, we shall refer to this framework as the information geometric framework~\cite{Amari85}.  

Within this framework, we formalize three elementary features of quantum phenomena, namely complementarity, global gauge invariance, and measurement simulability, detailed below.  These features can be understood as assertions about the physical world quite apart from the setting of the quantum formalism within which they are usually encountered~\cite{Goyal-QT1b}, and are sufficiently simple to be taken as primitives in the building up of quantum theory.  To these features, we add an information-theoretic principle, the principle of metric invariance.  From these ingredients, we reconstruct the principal features of the finite-dimensional quantum formalism, namely that pure states are represented by complex vectors, physical transformations are represented by unitary or antiunitary transformations, and the outcome probabilities~(and the corresponding output states) of measurements are given by the Born rule.   The present paper provides a streamlined derivation of the key parts of the finite-dimensional quantum formalism, focussing on the essential ideas.  The reader is referred to Refs.~\cite{Goyal-QT1b, Goyal-QT2b} for a more detailed discussion of the underlying ideas and methodology, as well as a derivation of the remainder of the finite-dimensional quantum formalism.

\section{Information Metric.}  We begin by giving a simple argument which shows how the information metric arises in a classical probabilistic setting from the concept of information.  Suppose that Alice has two coins,~$A$ and~$B$, characterized by the probability distributions~$\lvect{p} = (p_1, p_2)$ and~$\lvect{p}' = (p_1', p_2')$, respectively.  Suppose that she chooses coin~$A$, tosses it~$n$ times, and then sends the data to Bob, without disclosing to him which coin she chose.  If Bob knows~$\lvect{p}$ and~$\lvect{p}'$, how much information does the data provide him about which coin was tossed?  Intuitively, the more information the data provides, the more sharply the distributions are distinguished.

Using Bayes' theorem and Stirling's approximation for the case where~$n$ is large, on the assumption that coins~$A$ and~$B$ are \emph{a priori} equally likely to be chosen, one finds that
\begin{equation}
\frac{P_A}{P_B} =  \exp\left( n \sum_{i=1}^2 p_i \ln \frac{p_i}{p_i'} \right),
\end{equation}
where~$P_A$ is the probability that the tossed coin is~$A$ given the data, and likewise for~$P_B$~\footnote{See~\cite{Sivia96}, Chapter~4, for a discussion of such applications of Bayes' theorem}.
When the probability distributions are close, so that~$\lvect{p}' = \lvect{p} + d\lvect{p}$, the argument of the exponent can be expanded in the~$dp_i$ to give
\begin{equation}
\frac{P_A}{P_B} = \exp\left(2n\, ds^2\right), 
\end{equation}
where~$ds^2 = \frac{1}{4}\sum_i dp_i^2/p_i$ is the information metric.

Now, the information gained by Bob,~$\Delta I$, is the reduction in his uncertainty, and is therefore defined as
\begin{equation}
\Delta I \equiv U(1/2, 1/2) - U(P_A, P_B),
\end{equation}
with~$U$ being an entropy~(uncertainty) function such as the Shannon entropy.   But, since~$P_A + P_B = 1$ and~$P_A/P_B$ is determined by~$ds$, once~$U$ is selected,~$\Delta I$ is determined by~$ds$.  For example, if~$U$ is chosen to be the Shannon entropy~$U(\pi_1, \pi_2) = -\sum_i \pi_i \ln \pi_i$, one finds that
\begin{equation}
\Delta I =  \frac{1}{2} (n\, ds^2)^2.
\end{equation}
This result immediately generalizes to the case where~$\lvect{p}$ and~$\lvect{p}'$ are $M$-dimensional probability distributions~($M \ge 2$).   Hence, from an informational viewpoint, it is natural to endow the space of discrete probability distributions with the information metric.
 
Parenthetically, we remark that Wootters' statistical distance,~$\metricd_S(\lvect{p}, \lvect{p}') = \cos^{-1}\left(\sum_i \sqrt{p_i p'_i}\right)$, between the probability distributions~$\lvect{p}$ and~$\lvect{p}'$ is the minimum distance between~$\lvect{p}$ and~$\lvect{p}'$ with respect to the information metric~%
\footnote{This can most easily be seen by making a change of coordinates, so that~$q_i = \sqrt{p}_i$.  In terms of these coordinates, the information metric becomes Euclidean,~$ds^2 = \sum_i dq_i^2$, and the probability simplex becomes the positive orthant of the unit hypersphere in the $q$-space.  The minimum distance between the vectors~$\lvect{q} = (\sqrt{p}_1, \dots, \sqrt{p}_N)$ and~$\lvect{q}' = (\sqrt{p}'_1, \dots, \sqrt{p}'_N)$ is then simply~$\cos^{-1}(\lvect{q} . \lvect{q}')$ which is~$\metricd_S(\lvect{p}, \lvect{p}')$.%
}. %
We do not, however, make use of this result in what follows.

\section{Derivation}

\subsection{Construction of State Space.}
Measurement is idealized as a process that (i)~when performed upon some physical system, yields one of~$N$ possible \emph{outcomes,} with probabilities,~$p_1, \dots, p_N$, that are  determined by the state of the system immediately prior to the measurement, and (ii)~is reproducible, so that, upon immediate repetition of the measurement, the same outcome is obtained with certainty.

\subsubsection{Formalizing Complementarity.}

We take the first feature, complementarity, to consist of the general idea that, when a measurement is performed upon a system in some state, the measurement outcome only yields information about \emph{half} of the experimentally-accessible degrees of freedom of the state.   In the above classical probabilistic model of measurement, we can express this idea in a very simple way as follows:
\begin{itemize}
\item[]\emph{Postulate 1.}  \textbf{Complementarity}.  When measurement~$\mment{A}$ is performed, one of $2N$~possible \emph{events} occur, but they are not individually observed.  Outcome~$i$ is observed~($i=1, \dots, N$) whenever either event~$2i-1$ or event~$2i$ is realized.  The events~$1, \dots, 2N$ are assumed to occur with probabilities~$P_1, \dots, P_{2N}$, respectively, so that
\begin{equation} \label{eqn:p-P-relations}
p_i = P_{2i-1} + P_{2i},
\end{equation}
where~$p_i$ is the probability of outcome~$i$.
\end{itemize}
The~$P_q$~($q=1, \dots, 2N$) can be  summarized by the probability $n$-tuple~$\lvect{P} = (P_1, \dots, P_{2N})$.
As a result, of the~$2N-1$ degrees of freedom of~$\lvect{P}$, the measurement outcome only yields information about the~$p_i$, which constitute~$N-1$ degrees of freedom.   We shall shortly impose an additional constraint~(global gauge invariance) which implies that only~$2(N-1)$ of the~$2N-1$ degrees of freedom of~$\lvect{P}$ are physically relevant.  Hence, the measurement yields information about exactly one half of the experimentally-accessible degrees of freedom in~$\lvect{P}$.

Intuitively, performing the measurement brings about the realization of one of~$2N$ possible events but the observed outcomes \emph{coarse-grain} over these events:~when event~$2i-1$ or~$2i$ occurs, the measurement is~(for some reason to be investigated) unable to resolve the individual events, so that only outcome~$i$ is registered.   This is a novel hypothesis, which, at this point in the derivation, is recommended by its simplicity, and remains to be judged by its explanatory power~(namely its capacity to support a derivation of the quantum formalism)~\footnote{A similar hypothesis~(the `Knowledge Balance Principle') has been made in a recent toy model of quantum theory in order to give concrete expression to complementarity~\cite{Spekkens-toy-model}.  The insights provided by this toy model provides additional reason to explore the complementarity hypothesis given here.  See also Discussion section of Ref.~\cite{Goyal-QT1b}.}.

\subsubsection{Imposing the Information Metric.}

Next, we endow the space of probability distributions~$\lvect{P}$ with the information metric,~$ds^2 = \frac{1}{4} \sum_q dP_q^2/P_q$, where~$q=1, \dots, 2N$.    It is convenient to  define~$Q_q = \sqrt{P_q}$, where~$Q_q \in [0,1]$, since  the metric over the~$Q_q$ is then simply the Euclidean metric,~$ds^2  = dQ_1^2 + \dots  + dQ_{2N}^2$, so that~$\lvect{Q} = (Q_1, Q_2, \dots, Q_{2N})\transpose$ is a unit vector that lies on the positive orthant of the unit hypersphere~$\hypersphere$ is a $2N$-dimensional Euclidean space.

\subsubsection{Representing Physical Transformations.}

We now consider transformations of state space which represent physical transformations of the system.   We postulate that transformations of the state space, assumed one-to-one, preserve the metric over state space --- that is, the \emph{information distance},~$\metricd(\lvect{Q}, \lvect{Q}')$, between any pair of infinitesimally close states,~$\lvect{Q}, \lvect{Q}'$, where~$\metricd(\cdot)$ denotes distance with respect to the metric over state space, is preserved.   The essential idea here is that the discriminability of any pair of nearby states is a quantity that is intrinsic to this pair of states, and is therefore should remain invariant under reversible and deterministic transformations of the system~\footnote{Additionally, using the Measurement Simulability postulate given in Sec.~\ref{sec:representation-of-measurements}, this postulate can be grounded in the idea that the information distance  between any pair of nearby states should be the same irrespective of the measurement from whose perspective the states are observed.}.

Now, if one takes the~$\lvect{Q}$ themselves as the state space of the system, one immediately finds that continuous one-to-one transformations of the state space that preserve the information metric are not possible.  A simple way to allow the existence of such transformations is to take the entire unit hypersphere,~$S^{2N-1}$, as the state space of the system.   That is, we take the state of the system as been given by a unit vector~$\lvect{Q} = (Q_1, Q_2, \dots, Q_{2N})\transpose$,  with~$Q_q \in [-1, 1]$, where the probabilities~$P_q$ are given by~$P_q = Q_q^2$.  From the information metric over the~$\lvect{P}$, it follows from the relation~$P_q = Q_q^2$ that the metric over the~$\lvect{Q}$ is Euclidean,
\begin{equation}
\label{eqn:Q-space-metric}
ds^2 = dQ_1^2 + dQ_2^2 + \dots + dQ_{2N}^2.
\end{equation}

We can summarize the above requirements as follows:
\begin{itemize}
\item[]\emph{Postulate 2.} \textbf{Metric Invariance.}   The state of the system is given by the unit vector~$\lvect{Q} = (Q_1, Q_2, \dots, Q_{2N})\transpose$,  with~$Q_q \in [-1, 1]$, where the probabilities~$P_q$ are given by~$P_q = Q_q^2$.  The metric over the~$\lvect{Q}$ is Euclidean,~$ds^2 = dQ_1^2 + dQ_2^2 + \dots + dQ_{2N}^2$, which any transformation,~$\map{M}$, of state space must preserve.
\end{itemize}
It follows from this postulate that~$\lvect{Q}$  lies on the unit hypersphere,~$\hypersphere$, in a $2N$-dimensional real Euclidean space. From the requirement of metric preservation, it follows that~$\map{M}$~is an orthogonal transformation of~$\hypersphere$, so that every transformation can be expressed as~$\lvect{Q}' = \rmatrix{M}\lvect{Q}$, where~$\rmatrix{M}$ is a~$2N$-dimensional real orthogonal matrix.

The above extension of the state space from the positive orthant of~$\hypersphere$ to the entire hypersphere is an assumption which, although formally rather natural, presently awaits a clear physical basis.

\subsection{Global Gauge Invariance.}

The second feature, \emph{global gauge invariance}, consists of the idea that one can find a representation of the state of a system such that, if one displaces a subset of the degrees of freedom of the state by the same amount, any physical predictions based on the state are left invariant.    To formalize  this feature, we begin by making a change of variables by expressing the state,~$\lvect{Q}$,  in terms of the probabilities~$p_1, p_2, \dots, p_N$, and $N$~additional real degrees of freedom,~$\theta_1, \theta_2, \dots, \theta_N$, so that, without loss of generality,
\begin{equation} \label{eqn:Q-p-var-relations}
\begin{aligned}
Q_{2i-1} &= \sqrt{p_i} \cos\theta_i\\
Q_{2i} &= \sqrt{p_i} \sin\theta_i.
\end{aligned}
\end{equation}
Only the~$\theta_i$ can be subject to displacement since a displacement involving any of the~$p_i$ would be experimentally detectable. Accordingly, we formalize the idea of global gauge invariance by requiring that~$\theta_i = \theta(\var_i)$, where~$\theta(\cdot)$ is an unknown, non-constant, differentiable function to be determined, and that the transformation~$\var_i  \rightarrow \var_i + \var_0$ for~$i=1, \dots, N$ brings about no predictive changes for any~$\var_0 \in \numberfield{R}$.   From this global gauge condition, we immediately draw the following postulate:
\begin{itemize}
\item[]  \emph{Postulate 3.} \textbf{Gauge Invariance.}
	The map~$\map{M}$ is such that,
    for any state~$\lvect{Q} \in \hypersphere$, the
    probabilities,~$p_1', p_2', \dots, p_N'$,
    of the outcomes of measurement~$\mment{A}$ performed upon a system
    in state~$\lvect{Q}'=\map{M}(\lvect{Q})$ are unaffected if, in any
    representation,~$(p_i; \var_i) $, of the state~$\lvect{Q}$,
    an arbitrary real constant,~$\var_0$, is added
    to each of the~$\var_i$. 
\end{itemize}
Additionally, we draw the the requirement that the measure,~$\mu(p_i;  \var_i)$, over~$p_1, \dots, p_N, \var_1, \dots, \var_N$ induced by the metric over~$\hypersphere$ is consistent with the global gauge condition. This requirement is necessary in order that probabilistic inference using the measure as a prior over state space is consistent with our physical knowledge of the system.  This requirement yields the following postulate:
\begin{itemize}
\item[] \emph{Postulate 4.} \textbf{Measure Invariance.}
            The measure~$\mu(p_i;  \var_i)$ induced by the metric over state space satisfies the condition~$\mu(p_1, \dots, p_N; \var_1, \dots, \var_N) = \mu(p_1, \dots, p_N; \var_1 + 	\var_0, \dots, \var_N + \var_0)$ for any~$\var_0$.
\end{itemize}

\subsubsection{Determining the function~$\theta(\cdot)$.}

From Eqs.~\eqref{eqn:p-P-relations},~\eqref{eqn:Q-space-metric}, and~\eqref{eqn:Q-p-var-relations},
\begin{equation}
ds^2  =\frac{1}{4} \sum_{i=1}^N \frac{dp_i^2}{p_i} +  \sum_{i=1}^N p_i \theta'^2(\var_i) \, d\var_i^2.
\end{equation}
The measure,~$\mu(p_i; \var_i)$, over~$(p_1, \dots, p_N; \var_1, \dots, \var_N)$ induced by this metric is proportional to the square-root of the determinant of the metric,  and marginalizes to give
\begin{equation} \label{eqn:mu-i}
\mu_i(\var_i)  = c | \theta'(\var_i)|
\end{equation}
as the measure over~$\var_i$, where~$c$ is a constant.

Now, from the Measure Invariance postulate, it follows by marginalization that the measure~$\mu_i(\var_i)$ satisfies the relation~$\mu_i(\var_i + \var_0) = \mu_i(\var_i)$ for all~$\var_0$, and is therefore independent of~$\var_i$.  Hence, from Eq.~\eqref{eqn:mu-i},~$\theta(\var) = a\var + b$, where~$a, b$ are constants, where~$a \neq 0$ since, by assumption, the function~$\theta(\cdot)$ is not constant. We can therefore write
\begin{equation} \label{eqn:defn-of-lvect-Q-2}
\lvect{Q}   = (\sqrt{p_1} \cos\theta_1, \sqrt{p_1} \sin\theta_1, \ldots,
             \sqrt{p_N} \sin\theta_N).
\end{equation}

\subsubsection{Implementing Gauge Invariance, and the emergence of Complex Vector Space.}

From Eq.~\eqref{eqn:defn-of-lvect-Q-2}, the Gauge Invariance postulate, and the relation~$\theta_i = a\var_i + b$ given above, one can show that~$\rmatrix{M}$ is restricted to one of two \emph{types}:~$\rmatrix{M}$~has the general form
\begin{equation} \rmatrix{M} =
\begin{pmatrix} \label{eqn:restricted-rotation}
     T^{(11)} &  T^{(12)} & \dots &
        T^{(1N)} \\
     T^{(21)} &  T^{(22)} & \dots &
         T^{(2N)} \\
    \hdotsfor[2]{4}                 \\
     T^{(N1)} &  T^{(N2)} & \dots &
         T^{(NN)}
\end{pmatrix},
\end{equation}
where~$T^{(ij)}$ has the form
\[ T^{(ij)} =  \alpha_{ij}
        \begin{pmatrix}
          \cos \varphi_{ij}  & -\sin\varphi_{ij} \\
          \sin \varphi_{ij}  &  \cos\varphi_{ij}
        \end{pmatrix}
          \begin{pmatrix}  1 & 0 \\  0 & -1 \end{pmatrix}^{\beta},
\]
and where either~$\beta=0$~(type~1), in which case~$T^{(ij)}$ is a scale-rotation matrix, or~$\beta=1$~(type~2), in which case~$T^{(ij)}$ is a scale-rotation-reflection matrix,
with scale factor~$\alpha_{ij}$ and rotation angle~$\varphi_{ij}$ in either case~\footnote{See~\cite{Goyal-QT1b}, Sec.~V\,B\,2.}.   

Now, the state~$\lvect{Q}$ can be faithfully represented by the complex unit vector
\begin{equation} \label{eqn:def-of-v}
\begin{aligned}
\cvect{v} &\equiv (Q_1 + iQ_2, \dots, Q_{2N-1} + iQ_{2N})^\text{\textsf{T}} \\
		&= (\sqrt{p_1} e^{i\theta_1}, \dots, \sqrt{p_N} e^{i\theta_N})^\text{\textsf{T}},
\end{aligned}
\end{equation}
and, remarkably, one can then show that every transformation~$\rmatrix{M}$ of type~1 corresponds one-to-one with the set of unitary transformations of~$\cvect{v}$, and that every transformation~$\rmatrix{M}$ of type~2 corresponds one-to-one with the set of antiunitary transformations of~$\cvect{v}$.   In particular, on the assumption that a parameterized transformation that represents a continuous physical transformation must reduce to the identity for some value of the parameters, it follows that a continuous transformation must be represented by unitary transformations. 

\subsection{Representation of Measurements.}
\label{sec:representation-of-measurements} 

\begin{figure}[!h]
\begin{centering}
\includegraphics[width=3.25in]{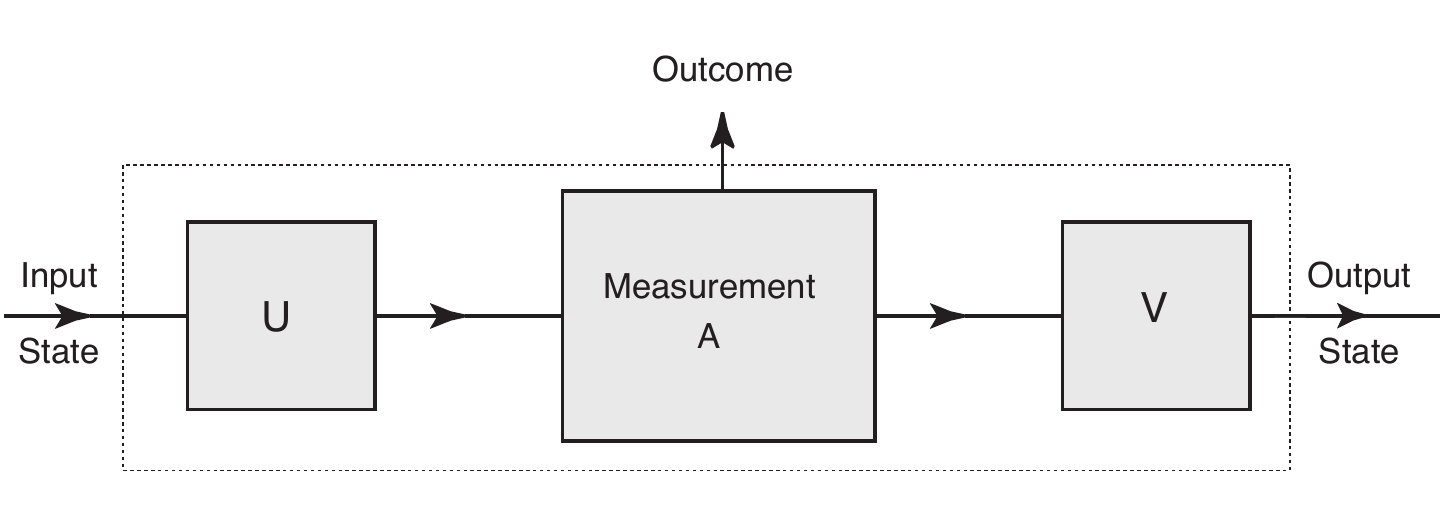}
\caption{\label{fig:representation-of-measurement} 
Simulation of measurement~$\mment{A}'$ in terms of measurement~$\mment{A}$.}
\end{centering}
\end{figure}

The third feature, \emph{measurement simulability}, can be stated as follows:
\begin{itemize}
\item[]\emph{Postulate 5.} \textbf{Measurement Simulability.}   Any reproducible measurement,~$\mment{A}'$, describable in the formalism can, insofar as its outcome probabilities and associated output states are concerned, be simulated by an arrangement consisting of measurement~$\mment{A}$ flanked by suitable interactions with the system.  
\end{itemize}
Given the results derived above, this postulate immediately implies that~$\mment{A}'$ can be simulated by the arrangement shown in Fig.~\ref{fig:representation-of-measurement}, where~$\cmatrix{U}$ and~$\cmatrix{V}$ are unitary transformations representing the interactions with the system.

The reproducibility of measurement~$\mment{A}$ implies that the state of a system immediately after~$\mment{A}$ has yielded outcome~$i$ is given by~$\cvect{v}_i = ( 0,\dots, e^{i\phi_i}, \dots, 0)^{\text{\textsf{T}}}$, where~$\phi_i$ is undetermined.  Hence, the input state~$\cvect{v}_i' = \cmatrix{U}^{-1} \cvect{v}_i$ will yield outcome~$i$.  In order that the arrangement behave like a reproducible measurement, the output state must be~$\cvect{v}_i'$ up to an overall phase, so that it suffices to choose~$\cmatrix{V}\cvect{v}_i = \cvect{v}_i'$ for~$i=1, \dots, N$, which implies that~$\cmatrix{V} = \cmatrix{U}^{-1}$.

Since the~$\cvect{v}_i$ form an orthonormal basis, it follows from~$\cvect{v}_i' = \cmatrix{U}^{-1} \cvect{v}_i$ that the~$\cvect{v}'_i$ also form an orthonormal basis.  Therefore, any state~$\cvect{v}$ can be expanded as~$\sum_i c_i' \cvect{v}'_i$, where~$c_i = {\cvect{v}'_i}^\dag \cvect{v}$.  With the input state~$\cvect{v}$, the state measured by measurement~$\mment{A}$ in the arrangement is~$\cmatrix{U}\cvect{v} = \sum_i c_i' \cvect{v}_i$.   From Eq.~\eqref{eqn:def-of-v}, the probabilities,~$p_1, \dots, p_N$, of the outcomes of measurement~$\mment{A}$ performed on state~$\cvect{v} = (v_1, \dots, v_N)$ are given by~$p_i = |v_i|^2$.  Therefore, in this case, the measurement yields outcome~$i$, together with output state~$\cvect{v}_i'$, with probability~$|c_i'|^2 = |{\cvect{v}'_i}^\dag \cvect{v}|^2$, which is the Born rule.

\section{Discussion}

The physical irrelevance of the overall phase of a pure state is usually regarded as being a minor mathematical feature of the quantum formalism of little physical importance.  From this standpoint, one of the most surprising finding in the derivation is that the global gauge condition~(which expresses in a more general way the physical irrelevance of the overall phase) is sufficiently strong as to transform a $2N$-dimensional real formalism~(where states are real unit vectors, and the transformations are the orthogonal transformations) into the familiar $N$-dimensional complex vector formalism of quantum theory~(where states are complex unit vectors, and the transformations are the unitary and antiunitary transformations).    In particular, the fact that the set of possible transformations one obtains is \emph{precisely} the set of all unitary and antiunitary transformations~(and neither more nor less) is not something that could, \emph{a priori}, have been reasonably anticipated.   

The derivation provides a number of other important insights into the structure of the quantum formalism.  From the perspective of the derivation, it is clear that the use of complex numbers in the quantum formalism is directly tied to the set of possible transformations of state space.  For example, if the set of \emph{all} orthogonal transformations were allowed, then the complex form of the formalism, whilst still possible to write down, would involve non-linear continuous transformations and would therefore not appear mathematically natural.     The derivation also suggests that information geometry is directly or indirectly responsible for many of its key mathematical features~(such as the importance of square-roots of probability, and the sinusoidal functions that appear in a quantum state), thereby providing significant new support for the hypothesis that information plays a fundamental role in determining the structure of quantum theory.

Finally, the derivation illuminates a previous partial reconstruction of quantum theory due to Stueckelberg~\cite{Stueckelberg60}.   Stueckelberg makes an assumption similar to the Complementarity postulate to arrive at the idea that the state of a system is given by a $2N$-dimensional probability distribution which can be written as a unit vector in a $2N$-dimensional `square-root of probability space', as we have done.  He then asserts that the allowable transformations of the state space are orthogonal transformations, and shows that, if the transformations are restricted by a superselection rule, then the set of restricted transformations is equivalent to the set of unitary transformations acting on a suitably-defined $N$-dimensional complex state space.   The present derivation shows that Stueckelberg's assertion that the allowable transformations are orthogonal transformations can be naturally accounted for in terms of the information metric over the probability simplex via the Metric Invariance postulate.    The derivation also shows that Stueckelberg's superselection rule can be replaced by the Global Gauge Invariance postulate.

\begin{acknowledgments}

I would like to thank Harvey Brown, Steve Flammia, Yiton Fu, Chris Fuchs, Lucien Hardy, Gerard 't~Hooft, Lane Hughston, and Lee Smolin for discussions and invaluable comments and suggestions.  I would also like to thank an anonymous reviewer for helpful suggestions.  Research at Perimeter Institute is supported in part by the Government of Canada through NSERC and by the Province of Ontario through MEDT. 

\end{acknowledgments}


\begin{thebibliography}{28}
\expandafter\ifx\csname natexlab\endcsname\relax\def\natexlab#1{#1}\fi
\expandafter\ifx\csname bibnamefont\endcsname\relax
  \def\bibnamefont#1{#1}\fi
\expandafter\ifx\csname bibfnamefont\endcsname\relax
  \def\bibfnamefont#1{#1}\fi
\expandafter\ifx\csname citenamefont\endcsname\relax
  \def\citenamefont#1{#1}\fi
\expandafter\ifx\csname url\endcsname\relax
  \def\url#1{\texttt{#1}}\fi
\expandafter\ifx\csname urlprefix\endcsname\relax\def\urlprefix{URL }\fi
\providecommand{\bibinfo}[2]{#2}
\providecommand{\eprint}[2][]{\url{#2}}

\bibitem[{\citenamefont{Fuchs}(2002)}]{Fuchs02}
\bibinfo{author}{\bibfnamefont{C.~A.} \bibnamefont{Fuchs}}
  (\bibinfo{year}{2002}), \eprint{quant-ph/0205039}.

\bibitem[{\citenamefont{Weinberg}(1989{\natexlab{a}})}]{Weinberg89a}
\bibinfo{author}{\bibfnamefont{S.}~\bibnamefont{Weinberg}},
  \bibinfo{journal}{Phys. Rev. Lett.} \textbf{\bibinfo{volume}{62}},
  \bibinfo{pages}{485} (\bibinfo{year}{1989}{\natexlab{a}}).

\bibitem[{\citenamefont{Weinberg}(1989{\natexlab{b}})}]{Weinberg89b}
\bibinfo{author}{\bibfnamefont{S.}~\bibnamefont{Weinberg}},
  \bibinfo{journal}{Ann. Phys. (N.Y.)} \textbf{\bibinfo{volume}{194}},
  \bibinfo{pages}{336} (\bibinfo{year}{1989}{\natexlab{b}}).

\bibitem[{\citenamefont{Herbert}(1982)}]{Herbert82}
\bibinfo{author}{\bibfnamefont{N.}~\bibnamefont{Herbert}},
  \bibinfo{journal}{Found. Phys.} \textbf{\bibinfo{volume}{12}},
  \bibinfo{pages}{1171} (\bibinfo{year}{1982}).

\bibitem[{\citenamefont{Isham}(2002)}]{Isham-quantum-and-reals}
\bibinfo{author}{\bibfnamefont{C.~J.} \bibnamefont{Isham}}
  (\bibinfo{year}{2002}), \eprint{quant-ph/0206090}.

\bibitem[{\citenamefont{Wheeler}(1989)}]{Wheeler89}
\bibinfo{author}{\bibfnamefont{J.~A.} \bibnamefont{Wheeler}}, in
  \emph{\bibinfo{booktitle}{Proceedings of the 3rd international symposium on
  the foundations of quantum mechanics, Tokyo}} (\bibinfo{year}{1989}).

\bibitem[{\citenamefont{Rovelli}(1996)}]{Rovelli96}
\bibinfo{author}{\bibfnamefont{C.}~\bibnamefont{Rovelli}},
  \bibinfo{journal}{Int. J. Theor. Phys.} \textbf{\bibinfo{volume}{35}},
  \bibinfo{pages}{1637} (\bibinfo{year}{1996}), \eprint{quant-ph/9609002v2}.

\bibitem[{\citenamefont{Popescu and Rohrlich}(1997)}]{Popescu-Rohrlich97}
\bibinfo{author}{\bibfnamefont{S.}~\bibnamefont{Popescu}} \bibnamefont{and}
  \bibinfo{author}{\bibfnamefont{D.}~\bibnamefont{Rohrlich}}, in
  \emph{\bibinfo{booktitle}{Causality and Locality in Modern Physics and
  Astronomy: Open Questions and Possible Solutions}} (\bibinfo{year}{1997}),
  \eprint{quant-ph/9709026}.

\bibitem[{\citenamefont{Zeilinger}(1999)}]{Zeilinger99}
\bibinfo{author}{\bibfnamefont{A.}~\bibnamefont{Zeilinger}},
  \bibinfo{journal}{Found. Phys.} \textbf{\bibinfo{volume}{29}},
  \bibinfo{pages}{631} (\bibinfo{year}{1999}).

\bibitem[{\citenamefont{Hardy}(2001)}]{Hardy01a}
\bibinfo{author}{\bibfnamefont{L.}~\bibnamefont{Hardy}} (\bibinfo{year}{2001}),
  \eprint{quant-ph/0101012}.

\bibitem[{\citenamefont{Wootters}(1980)}]{Wootters80}
\bibinfo{author}{\bibfnamefont{W.~K.} \bibnamefont{Wootters}}, Ph.D. thesis,
  \bibinfo{school}{University of Texas at Austin} (\bibinfo{year}{1980}).

\bibitem[{\citenamefont{Summhammer}(1994)}]{Summhammer94}
\bibinfo{author}{\bibfnamefont{J.}~\bibnamefont{Summhammer}},
  \bibinfo{journal}{Int. J. Theor. Phys.} \textbf{\bibinfo{volume}{33}},
  \bibinfo{pages}{171} (\bibinfo{year}{1994}), \eprint{quant-ph/9910039}.

\bibitem[{\citenamefont{Brukner and Zeilinger}(1999)}]{Brukner99}
\bibinfo{author}{\bibfnamefont{{\v{C}.}.}~\bibnamefont{Brukner}}
  \bibnamefont{and}
  \bibinfo{author}{\bibfnamefont{A.}~\bibnamefont{Zeilinger}},
  \bibinfo{journal}{Phys. Rev. Lett.} \textbf{\bibinfo{volume}{83}},
  \bibinfo{pages}{3354} (\bibinfo{year}{1999}).

\bibitem[{\citenamefont{\v{C}. Brukner and Zeilinger}(2002)}]{Brukner02b}
\bibinfo{author}{\bibnamefont{\v{C}. Brukner}} \bibnamefont{and}
  \bibinfo{author}{\bibfnamefont{A.}~\bibnamefont{Zeilinger}}, in
  \emph{\bibinfo{booktitle}{Time, Quantum, and Information}}, edited by
  \bibinfo{editor}{\bibfnamefont{L.}~\bibnamefont{Castell}} \bibnamefont{and}
  \bibinfo{editor}{\bibfnamefont{O.}~\bibnamefont{Ischebeck}}
  (\bibinfo{publisher}{Springer}, \bibinfo{year}{2002}),
  \eprint{quant-ph/0212084v1}.

\bibitem[{\citenamefont{Grinbaum}(2003)}]{Grinbaum03}
\bibinfo{author}{\bibfnamefont{A.}~\bibnamefont{Grinbaum}},
  \bibinfo{journal}{Int. J. Quant. Inf.} \textbf{\bibinfo{volume}{1}},
  \bibinfo{pages}{289} (\bibinfo{year}{2003}), \eprint{quant-ph/0306079}.

\bibitem[{\citenamefont{Grinbaum}(2004)}]{Grinbaum04}
\bibinfo{author}{\bibfnamefont{A.}~\bibnamefont{Grinbaum}}, Ph.D. thesis,
  \bibinfo{school}{Ecole Polytechnique, Paris} (\bibinfo{year}{2004}),
  \eprint{{quant-ph/0410071}}.

\bibitem[{\citenamefont{Caticha}(2000)}]{Caticha99b}
\bibinfo{author}{\bibfnamefont{A.}~\bibnamefont{Caticha}},
  \bibinfo{journal}{Found. Phys.} \textbf{\bibinfo{volume}{30}},
  \bibinfo{pages}{227} (\bibinfo{year}{2000}), \eprint{quant-ph/9810074v2}.

\bibitem[{\citenamefont{Clifton et~al.}(2003)\citenamefont{Clifton, Bub, and
  Halvorson}}]{Clifton-Bub-Halvorson03}
\bibinfo{author}{\bibfnamefont{R.}~\bibnamefont{Clifton}},
  \bibinfo{author}{\bibfnamefont{J.}~\bibnamefont{Bub}}, \bibnamefont{and}
  \bibinfo{author}{\bibfnamefont{H.}~\bibnamefont{Halvorson}},
  \bibinfo{journal}{Found. Phys.} \textbf{\bibinfo{volume}{33}},
  \bibinfo{pages}{1561} (\bibinfo{year}{2003}).

\bibitem[{\citenamefont{Wootters}(1981)}]{Wootters-statistical-distance}
\bibinfo{author}{\bibfnamefont{W.~K.} \bibnamefont{Wootters}},
  \bibinfo{journal}{Phys. Rev.~D} \textbf{\bibinfo{volume}{23}}
  (\bibinfo{year}{1981}).

\bibitem[{\citenamefont{Brody and Hughston}(1996)}]{Brody-Hughston96}
\bibinfo{author}{\bibfnamefont{D.~C.} \bibnamefont{Brody}} \bibnamefont{and}
  \bibinfo{author}{\bibfnamefont{L.~P.} \bibnamefont{Hughston}},
  \bibinfo{journal}{Phys. Rev. Lett.} \textbf{\bibinfo{volume}{77}},
  \bibinfo{pages}{2851} (\bibinfo{year}{1996}).

\bibitem[{\citenamefont{Brody and Hughston}(1998)}]{Brody-Hughston98}
\bibinfo{author}{\bibfnamefont{D.~C.} \bibnamefont{Brody}} \bibnamefont{and}
  \bibinfo{author}{\bibfnamefont{L.~P.} \bibnamefont{Hughston}},
  \bibinfo{journal}{Proc. R. Soc. Lond.~A} \textbf{\bibinfo{volume}{454}},
  \bibinfo{pages}{2445} (\bibinfo{year}{1998}).

\bibitem[{\citenamefont{Mehrafarin}(2005)}]{Mehrafarin05}
\bibinfo{author}{\bibfnamefont{M.}~\bibnamefont{Mehrafarin}},
  \bibinfo{journal}{Int. J. Theor. Phys.} \textbf{\bibinfo{volume}{44}},
  \bibinfo{pages}{429} (\bibinfo{year}{2005}), \eprint{quant-ph/0402153}.

\bibitem[{\citenamefont{Amari}(1985)}]{Amari85}
\bibinfo{author}{\bibfnamefont{S.}~\bibnamefont{Amari}},
  \emph{\bibinfo{title}{Differential-geometrical methods in statistics}}
  (\bibinfo{publisher}{Springer-Verlag}, \bibinfo{year}{1985}).

\bibitem[{\citenamefont{Goyal}(2008{\natexlab{a}})}]{Goyal-QT1b}
\bibinfo{author}{\bibfnamefont{P.}~\bibnamefont{Goyal}}
  (\bibinfo{year}{2008}{\natexlab{a}}), \eprint{arXiv:0805.2761v1}.

\bibitem[{\citenamefont{Goyal}(2008{\natexlab{b}})}]{Goyal-QT2b}
\bibinfo{author}{\bibfnamefont{P.}~\bibnamefont{Goyal}}
  (\bibinfo{year}{2008}{\natexlab{b}}), \eprint{arXiv:0805.2765v1}.

\bibitem[{\citenamefont{Stueckelberg}(1960)}]{Stueckelberg60}
\bibinfo{author}{\bibfnamefont{E.~C.~G.} \bibnamefont{Stueckelberg}},
  \bibinfo{journal}{Helv. Phys. Acta.} \textbf{\bibinfo{volume}{33}},
  \bibinfo{pages}{727} (\bibinfo{year}{1960}).

\bibitem[{\citenamefont{Sivia}(1996)}]{Sivia96}
\bibinfo{author}{\bibfnamefont{D.~S.} \bibnamefont{Sivia}},
  \emph{\bibinfo{title}{Data Analysis: A Bayesian Tutorial}}
  (\bibinfo{publisher}{Oxford Science Publications}, \bibinfo{year}{1996}).

\bibitem[{\citenamefont{Spekkens}(2007)}]{Spekkens-toy-model}
\bibinfo{author}{\bibfnamefont{R.~W.} \bibnamefont{Spekkens}},
  \bibinfo{journal}{Phys. Rev.~A} \textbf{\bibinfo{volume}{75}},
  \bibinfo{pages}{032110} (\bibinfo{year}{2007}).

\end{thebibliography}
\end{document}